\documentclass[12pt]{article}
\usepackage[cp1251]{inputenc}
\usepackage{amsmath}
\usepackage{amsfonts}
\usepackage{amssymb}
\def\pa{\partial}                       
\def\beq{\begin{eqnarray}}    
\def\eeq{\end{eqnarray}}      

\begin{document}
\date{}
\begin{center}
{\Large\textbf{On interactions of massless integer high spin and scalar
 fields }}

\vspace{18mm}

{\large
P.M. Lavrov$^{(a, b)} \footnote{E-mail: lavrov@tspu.edu.ru}$
}

\vspace{8mm}

\noindent  ${{}^{(a)}} ${\em
Center of Theoretical Physics, \\
Tomsk State Pedagogical University,\\
Kievskaya St.\ 60, 634061 Tomsk, Russia}

\noindent  ${{}^{(b)}} ${\em
National Research Tomsk State  University,\\
Lenin Av.\ 36, 634050 Tomsk, Russia}

\vspace{20mm}

\begin{abstract}
\noindent

We apply an unconstrained formulation of bosonic higher spin fields to study
interactions of these fields with a bosonic field using new method for the
deformation procedure. It is proved that local vertices of any order containing
one higher spin $s$ field and arbitrary number of scalar fields and being invariant
under original gauge transformations are described with the help of one local totally
symmetric $(s-2)$-rank tensor of scalar fields. This tensor is explicitly constructed
for particular  cases related to cubic vertices
for spin $s$ and vertices of an arbitrary order  for spin $s=4$.

\end{abstract}

\end{center}

\vfill

\\

\newpage

\section{Introduction}

Over the past forty years, the construction of interacting higher spin fields
has attracted increasing interest due to important obtained  results
 and numerous unsolved problems that arise
in this sphere of activity \cite{BBvD-1,BBvD-2,FV1,FV2}. Main results and
understanding  have been achieved
in studies of  cubic vertices
\cite{FIPT,BJM,MMR1,Zinov,MMR2,FMM,ZinKh20,BKTM21,BR,BKS22}
using different approaches
(deformation procedure in the BV formalism \cite{BH,H},
BRST construction \cite{BPT}, light-cone formalism \cite{Mets})
\!\footnote{Recently a supersymmetric extension in constructions of cubic vertices
\cite{BGatesK-1,BGatesK-2,Kout} and
description of related aspects \cite{KKvU,BGatesK-3} have been proposed.}.
Constructions of quartic vertices \cite{FT,Pol,Tar,DT,FKM19,JT19,KMP} require more efforts
and lead to the non-locality problem in the high spin theory.
A widely used method for constructing interaction vertices in a deformation
scheme embedded in the BV formalism \cite{BH,H} is the Noether procedure
(see, for instance, \cite{Zinov,MMR2,KMP} and references therein).

Recently,
the new method based on systematic use of anticanonical transformations
in the BV-formalism \cite{BV,BV1} to obtain a new approach for solutions to
the deformation procedure has been proposed \cite{BL-1,BL-2,L-2022}.
In fact, this method opens a new way in constructions of gauge models with
interactions between fields. Among achievements of new method there are models
containing quartic and quintic vertices for massless spin 3,
 massive vector and scalar fields
\cite{L-ejpc-s3,L-22-2,L-s5} as well as cubic vertices for massless spin 4,
 massive vector and scalar fields \cite{L-s4-3}. All these models are described in an exact
 and closed form in terms of local functionals of finite orders in fields which are invariant
 under original gauge transformations.

In the present paper, we are going to extend results of papers
\cite{L-ejpc-s3,L-22-2,L-s5,L-s4-3} to the case of massless integer spin $s$ fields
using general statements of the new method.

The paper is organized as follows. In section 2, an unconstrained formulation of
initial action for massless integer spin and scalar fields
is introduced.
In section 3, general conditions of anticanonical transformations leading to
gauge invariance of local part of deformed action are derived.
Section 4  is devoted to eliminating all auxiliary fields to obtain constrained
formulation for interactions of massless spin $s$ field with arbitrary number of scalar
fields. In section 5, two examples of vertices for model under consideration
are presented.
In section 6, concluding remarks are given.

The DeWitt's condensed notations
\cite{DeWitt} are systematically used. The right
functional derivatives are marked by special symbols $"\leftarrow"$.
Arguments of any functional
are enclosed in square brackets $[\;]$, and arguments of any
function are enclosed in parentheses, $(\;)$.

\section{Unconstrained action}
\noindent
The new method to the deformation procedure is formulated within the BV-formalism,
which operates with gauge theories of unconstrained fields. Massless integer high spin
fields in the Fronsdal theory \cite{Fronsdal-1} with spin $s> 3$ are satisfied
to double traceless conditions. For these cases,  direct application of the new method
ia not possible.
Fortunately, there exists an unconstrained formulation for massless integer
high spin fields \cite{BGK} that allows us to apply the new method.
The action for such fields
within the unconstrained formulation has the form
\beq
&&S_0[\varphi, F, \alpha, \lambda_{(2)}, \lambda_{(4)}]
=\int dx\Big[\varphi _{\mu _{1}\cdots \mu
_{s}}\Box\varphi ^{\mu _{1}\cdots \mu _{s}}-s(s-1)
F_{\mu _{1}\cdots \mu _{s-2}}\Box F^{\mu _{1}\cdots \mu _{s-2}}+
\notag \\
&&\qquad\qquad\qquad\qquad +s\Lambda _{\mu _{1}\mu _{2}
\cdots \mu _{s-1}}\Lambda ^{\mu _{1}\mu
_{2}\cdots \mu _{s-1}}+  \notag \\
&&\qquad\qquad\qquad\qquad +\lambda _{(2)}^{\mu _{1}\cdots
\mu _{s-2}}\left( \eta ^{\mu \nu
}\varphi _{\mu \nu \mu _{1}\cdots \mu _{s-2}}-2F_{\mu _{1}\cdots \mu
_{s-2}}-(s-2)\partial _{\mu _{1}}\alpha _{\mu _{2}\cdots \mu _{s-2}}\right] +
\notag \\
&&\qquad\qquad\qquad\qquad +\lambda _{(4)}^{\mu _{1}\cdots \mu _{s-4}}
\left( \eta ^{\mu \nu }F_{\mu
\nu \mu _{1}\cdots \mu _{s-4}}-\partial ^{\mu }\alpha _{\mu \mu _{1}\cdots
\mu _{s-4}}\right)\Big] ,
\label{b1}.
\eeq
where the notation\footnote{The symbol
$(\cdots)$ means the cycle permutation of indexes involved.}
\beq
\label{b2}
\Lambda _{\mu _{1}\mu _{2}\cdots \mu _{s-1}}=
\pa^{\mu}\varphi_{\mu\mu _{1}\mu _{2}\cdots \mu _{s-1}}-
\pa_{(\mu_1}F_{\mu _{2}\mu _{2}\cdots \mu _{s-1})}
\eeq
is used. In Eq. (\ref{b1})
$\varphi ^{\mu _{1}\cdots \mu _{s}}=
\varphi ^{\mu _{1}\cdots \mu _{s}}(x)$ is completely
symmetric  s-rank tensor,
$\Box$ is the D'Alembertian, $\Box=\pa_{\mu}\pa^{\mu}$,  $\eta_{\mu\nu}$ is
the metric tensor of flat Minkowski space of the dimension $d$, and
$F^{\mu _{1}\cdots \mu _{s-2}}$,
$\alpha ^{\mu _{1}\cdots \mu _{s-3}}$,
$\lambda _{(2)}^{\mu _{1}\cdots \mu _{s-2}}$,
$\lambda _{(4)}^{\mu _{1}\cdots \mu _{s-4}}$ are auxiliary
fields of corresponding ranks coinciding with number of indices.

We are going to construct  interactions
of massless integer spin $s$ fields with a real scalar field $\phi=\phi(x)$.
We assume that the initial action has the form
\beq
\label{c1}
S_0[A]=S_0[\varphi, F, \alpha, \lambda_{(2)},
\lambda_{(4)}]+S_0[\phi],
\eeq
where  $S_0[\phi]$ is the action of a
free real scalar field,
\beq \label{c1a} S_0[\phi]=\int
dx\frac{1}{2}\big[ \pa_{\mu}\phi\;\pa^{\mu}\phi-m^2\phi^2\big],
\eeq
and $A=\{A^i\}$ denotes collection of all fields of the model,
\beq
A^i=(\varphi^{\mu _{1}\cdots \mu _{s}}, \phi,\; F^{\mu _{1}\cdots \mu _{s-2}},\;
\alpha^{\mu _{1}\cdots \mu _{s-3}},\;
\lambda_{(2)}^{\mu _{1}\cdots \mu _{s-2}},\;
\lambda_{(4)}^{\mu _{1}\cdots \mu _{s-4}}).
\eeq
The action (\ref{c1})
is invariant under the gauge transformations
\beq
\nonumber
&&\delta\varphi^{\mu _{1}\cdots \mu _{s}}=
\pa^{(\mu_1}\xi^{\mu _{2}\cdots \mu _{s})},
\quad \delta F^{\mu _{1}\cdots \mu _{s-2}}=
\pa_{\lambda}\xi^{\lambda\mu _{1}\cdots \mu _{s_2}},\quad
\delta\alpha^{\mu _{1}\cdots \mu _{s-3}}=
\eta_{\lambda\nu}\xi^{\lambda\nu\mu _{1}\cdots \mu _{s-3}},\\
\label{c2}
&&
\delta\lambda_{(2)}^{\mu _{1}\cdots \mu _{s-2}}=0,
\quad \delta\lambda_{(4)}^{\mu _{1}\cdots \mu _{s-4}}=0,\quad
\delta \phi=0,
\eeq
where $\xi^{\mu _{1}\cdots \mu _{s-1}}$ is arbitrary totally symmetric $(s-1)$-rank tensor.
Algebra of gauge transformations is Abelian.

\section{Gauge-invariant deformations}

Now, we consider possible deformations of initial action using the procedure
which is ruled out
by the generating functions $h^i(A)$ \cite{BL-1}.
 Here, we restrict ourself by the case of
anticanonical transformations acting effectively in the sector of fields
$\varphi^{\mu_1\cdots\mu_{s}}$ and $F^{\mu_1\cdots\mu_{s-2}}$ of the initial theory.
It means the following structure of
generating functions
$h^i(A)=(h_{(\varphi)}^{\mu_1\cdots\mu_{s}}(\phi), 0,
h_{(F)}^{\mu_1\cdots\mu_{s-2}}(\phi), 0,0,0)$.
In construction of suitable generating
functions
$h_{(\varphi)}^{\mu_1\cdots\mu_{s}}=h_{(\varphi)}^{\mu_1\cdots\mu_{s}}(\phi)$
and $h_{(F)}^{\mu_1\cdots\mu_{s_2}}=h_{(F)}^{\mu_1\cdots\mu_{s}}(\phi)$, we have to take
into account the dimensions of quantities involved
in the initial action $S_0[A]$ (\ref{c1}),
\beq
\nonumber
&&{\rm dim}(\varphi^{\mu_1\cdots\mu_{s}})=
{\rm dim}(F^{\mu_1\cdots\mu_{s-2}})={\rm dim}(\phi)=\frac{d-2}{2},\\
\nonumber
&&{\rm dim}(\lambda_{(2)}^{\mu _{1}\cdots \mu _{s-2}})=
{\rm dim}(\lambda_{(4)}^{\mu _{1}\cdots \mu _{s-4}})=\frac{d+2}{2},\quad
{\rm dim}(\alpha^{\mu _{1}\cdots \mu _{s-3}})=\frac{d-4}{2},\\
\label{c3}
&&
{\rm dim}(\xi^{\mu_1\cdots\mu_{s-2}})=\frac{d-4}{2},\quad {\rm dim}(\pa_{\mu})=1, \;\;
{\rm dim}(\Box)=2.\quad {\rm dim}(\eta_{\mu\nu})=0 .
\eeq
The generating functions $h_{(\varphi)}^{\mu_1\cdots\mu_{s}}$ and
$h_{(F)}^{\mu_1\cdots\mu_{s-2}}$
should be symmetric and non-local  with
the dimension equals to $-(d+2)/2$.
The non-locality will be achieved by using the operator
$1/\Box$.
Therefore, the following presentation of generating functions of special anticanonical
transformations,
\beq
\label{c4}
h_{(\varphi)}^{\mu_{1}\cdots \mu_{s}}=
 \frac{1}{\Box}K^{\mu_{1}\cdots \mu_{s}},\quad
h_{(F)}^{\mu_{1}\cdots \mu_{s-2}}= \frac{1}{\Box}N^{\mu_{1}\cdots \mu_{s-2}}
\eeq
is used. Here $K^{\mu_{1}\cdots \mu_{s}}$ and $N^{\mu_{1}\cdots \mu_{s-2}}$
are assumed to be local
functions of fields $\phi$.

The deformed action has the following explicit and closed form
\beq
\label{c6}
\widetilde{S}[A] = S_0[A]+S_{int}[A],
\eeq
where
\beq
\nonumber
&&S_{int}[A]=\int dx \Big[2\varphi_{\mu_{1}\cdots \mu_{s}}
K^{\mu_{1}\cdots \mu_{s}}-
2s(s-1)F_{\mu_{1}\cdots \mu_{s-2}}N^{\mu_{1}\cdots \mu_{s-2}}+\\
\nonumber
&&\qquad\qquad+2s
\Lambda_{\mu_{1}\cdots \mu_{s=1}}\frac{1}{\Box}\big(\pa_{\mu}K^{\mu\mu_{1}\cdots \mu_{s-1}}-
2\pa^{(\mu_1}N^{\mu_{1}\cdots \mu_{s-1})}\big)+\\
\nonumber
&&\qquad\qquad+\lambda_{(2)}^{\mu_{1}\cdots \mu_{s-2}}\frac{1}{\Box}
\big(\eta^{\mu\nu}
K_{\mu\nu\mu_{1}\cdots \mu_{s-2}}-
2N_{\mu_{1}\cdots \mu_{s}}\big)+\lambda_{(4)}^{\mu_{1}\cdots \mu_{s-4}}
\frac{1}{\Box}\eta^{\mu\nu}N_{\mu\nu\mu_{1}\cdots \mu_{s-4}}+\\
\nonumber
&&\qquad\qquad+
K_{\mu_{1}\cdots \mu_{s}}\frac{1}{\Box}K^{\mu_{1}\cdots \mu_{s}}-
s(s-1)N_{\mu_{1}\cdots \mu_{s-2}}\frac{1}{\Box}N^{\mu_{1}\cdots \mu_{s-2}}+\\
\label{c7}
&&\qquad\qquad+s
\frac{1}{\Box}\big(\pa^{\mu}K_{\mu\mu_{1}\cdots \mu_{s-1}}\!-\!
2\pa_{(\mu_1}N_{\mu_{1}\cdots \mu_{s-1})}\big)
\frac{1}{\Box}\big(\pa_{\nu}K^{\nu\mu_{1}\cdots \mu_{s-1}}\!-\!
2\pa^{(\mu_1}N^{\mu_{1}\cdots \mu_{s-1})}\big)\Big].
\eeq

For theories with Abelian gauge algebra the deformation of gauge symmetry
is defined by the matrix $(M^{(-1)})^i_{\;j}(A)$ being inverse
to
$M^i_{\;j}(A)=\delta^i_{\;j}+h^i(A)\overleftarrow{\pa}_{\!A^j}$.
In the case under consideration the matrix $M^i_{\;j}(A)$ has
 the following structure in the sector of fields $\varphi, \phi, F$,
\beq
\label{c9}
\left(\begin{array}{ccc}
E^{\mu_{1}\cdots \mu_{s}}_{\nu_{1}\cdots \nu_{s}}&
h_{(\varphi)}^{\mu_{1}\cdots \mu_{s}}(\phi)\overleftarrow{\pa}_{\!\phi}& 0\\
0 & 1 & 0\\
0& h_{(F)}^{\mu_{1}\cdots \mu_{s-2}}(\phi)\overleftarrow{\pa}_{\!\phi}&
E^{\mu_{1}\cdots \mu_{s-2}}_{\nu_{1}\cdots \nu_{s-2}} \\
\end{array}\right).
\eeq
From Eq. (\ref{c9}) it follows that the matrix $(M^{(-1)})^i_{\;j}(A)$
can be found explicitly and
in the sector of fields $\varphi, \phi, F$ reads
\beq
\label{c10}
\left(\begin{array}{ccc}
E^{\mu_{1}\cdots \mu_{s}}_{\nu_{1}\cdots \nu_{s}}&-
h_{(\varphi)}^{\mu_{1}\cdots \mu_{s}}(\phi)\overleftarrow{\pa}_{\!\phi}& 0\\
0 & 1 & 0\\
0& - h_{(F)}^{\mu_{1}\cdots \mu_{s-2}}(\phi)\overleftarrow{\pa}_{\!\phi}&
E^{\mu_{1}\cdots \mu_{s-2}}_{\nu_{1}\cdots \nu_{s-2}} \\
\end{array}\right) .
\eeq
Here,  $E^{\mu_{1}\cdots \mu_{s}}_{\nu_{1}\cdots \nu_{s}}$
are elements of the unit matrix in the space of
symmetric  $s$-rank tensors. In particular, the presentation (\ref{c10}) means
that in process of deformation of the initial action the gauge transformations
(\ref{c2}) do not deform. In turn, the deformed action
\beq
\label{c11}
\delta\widetilde{S}[A]=0
\eeq
should be  invariant under original gauge transformations (\ref{c2}). As a result,
for the model under consideration we found explicit description of deformed
action and gauge symmetry.

Consider now the local part of action $S_{int}[A]$ (\ref{c7}),
\beq
\label{c12}
S_{1\; loc}[A]=2\int dx \Big[\varphi_{\mu_{1}\cdots \mu_{s}}
K^{\mu_{1}\cdots \mu_{s}}-
s(s-1)F_{\mu_{1}\cdots \mu_{s-2}}N^{\mu_{1}\cdots \mu_{s-2}}\Big].
\eeq
In fact, the action $S_{1\; loc}[A]$ depends on fields $\varphi, \phi, F$ only.
Due to the locality of original gauge symmetry and gauge invariance of the deformed action
(\ref{c11}), the action $S_{1\; loc}[A]$ should be
invariant under original gauge transformations as well,
\beq
\label{c13}
\delta S_{1\; loc}[A]=0.
\eeq
Taking into account the gauge transformations of fields $\varphi, \phi, F$, the equation
(\ref{c13}) is equivalent to
\beq
\label{c14}
\int dx \xi_{\mu_{2}\cdots \mu_{s}}\Big[\pa_{\mu_1}
K^{\mu_1\mu_{2}\cdots \mu_{s}}-(s-1)\pa^{\mu_2}N^{\mu_{3}\cdots \mu_{s}}\Big]=0.
\eeq
The structure of the last equation allows us to suppose that
$K^{\mu_1\mu_{2}\cdots \mu_{s}}$ should be proportional at least
two partial derivatives,
\beq
\label{c15}
K^{\mu_1\mu_{2}\cdots \mu_{s}}=\pa^{(\mu_1}\pa^{\mu_2}R^{\mu_{3}\cdots \mu_{s})}
\eeq
where $R^{\mu_{1}\cdots \mu_{s - 2}}$  is a local totally symmetric (s-2)-rank tensor.
Then, the equation (\ref{c14}) rewrites
\beq
\label{c16}
\int dx \xi_{\mu_{2}\cdots \mu_{s}}\pa^{\mu_2}
\Big[2\Box R^{\mu_{3}\cdots \mu_{s}}+
(s-2)\pa_{\rho}\pa^{\mu_3}R^{\mu_{4}\cdots \mu_{s}\rho}
-(s-1)N^{\mu_{3}\cdots \mu_{s}}\Big]=0.
\eeq
Choosing the generating function $N^{\mu_{1}\cdots \mu_{s-2}}$ in the form
\beq
\label{c16a}
N^{\mu_{1}\cdots \mu_{s-2}}=\frac{2}{s-1}\Box R^{\mu_{1}\cdots \mu_{s-2}}+
\frac{1}{s-1}\pa_{\rho}\pa^{(\mu_1}R^{\mu_{2}\cdots \mu_{s-2})\rho},
\eeq
the equation (\ref{c14}) describing the gauge invariance of local part of the deformed
action will be satisfied. Notice that the action $S_{loc}[A]$
\beq
\label{c17}
S_{loc}[A]=S_0[A]+S_{1\;loc}[\varphi,\phi,F],
\eeq
where
\beq
\nonumber
&&S_{1\;loc}[\varphi,\phi,F]=2s\int dx\Big[\varphi_{\mu_{1}\mu_2\cdots \mu_{s}}
\pa^{\mu_1}\pa^{\mu_2}R^{\mu_{3}\cdots \mu_{s}}(\phi)-
2F_{\mu_{1}\cdots \mu_{s-2}}\Box R^{\mu_{1}\cdots \mu_{s-2}}(\phi)-\\
\label{c18}
&&\qquad\qquad\qquad\qquad\qquad-
(s-2)F_{\mu_{1}\mu_2\cdots \mu_{s-2}}
\pa_{\rho}\pa^{\mu_1}R^{\mu_{2}\cdots \mu_{s-2}\rho}(\phi)
\Big],
\eeq
describes in exact and closed form of a local gauge invariant model containing (among others)
vertices
with one massless spin $s$ field and any number of scalar fields if the anticanonical
transformations (\ref{c4}) subject to the conditions (\ref{c15}), (\ref{c16a}).
For any choice of local
totally symmetric $(s-2)$-rank tensor
$R^{\mu_{1}\cdots \mu_{s-2}}(\phi)$ the local action (\ref{c18})
\beq
\nonumber
&&
S_{1\;loc}[\varphi,\phi,F]=2s\int dx\Big[\pa^{\mu_1}\pa^{\mu_2}
\varphi_{\mu_{1}\mu_2\mu_3\cdots \mu_{s}}
-
2\Box F_{\mu_{3}\cdots \mu_{s}}-\\
\label{c19}
&&\qquad\qquad\qquad\qquad\qquad-
(s-2)\pa_{\mu_3}\pa^{\sigma}F_{\sigma\mu_{4}\cdots \mu_{s}}
\Big]R^{\mu_{3}\cdots \mu_{s}}(\phi),
\eeq
is invariant
\beq
\label{c20}
\delta S_{1\;loc}[\varphi,\phi,F]=0,
\eeq
under the following gauge transformations
\beq
\label{c21}
\delta\varphi_{\mu_{1}\mu_2\cdots \mu_{s}}=
\pa_{(\mu_1}\xi_{\mu_{2}\cdots \mu_{s})}, \quad
\delta F_{\mu_{1}\mu_2\cdots \mu_{s-2}}=
\pa^{\mu}\xi_{\mu\mu_{1}\mu_2\cdots \mu_{s-2}}, \quad \delta\phi=0,
\eeq
where $\xi_{\mu_{1}\cdots \mu_{s-1}}$ are arbitrary functions of space-time coordinates.
Moreover, we prove that all arbitrariness
in special anticanonical transformations
leading in the initial model to the existence of local part of the deformed action
invariant under original
gauge transformations is described by one generating function
$R^{\mu_{1}\cdots \mu_{s-2}}(\phi)$.

\section{Eliminating auxiliary fields}

Usually in high spin theory interactions between  fields
which include at least one field with
spin $s>2$ are searched with the help of Noether procedure using fields subjected to
double traceless conditions. Main concrete results  in this direction are  related to
cubic vertices obtained in linear approximation with respect to the deformation parameter
(see, for example, \cite{FIPT,BJM,MMR1,MR,MM}).

The bridge between the standard approach and new method exists if one eliminates all auxiliary
fields appearing in the unconstrained formulation with the help of the equations of motion for
the initial action. They are
\beq
\nonumber
&&\eta^{\mu\nu}\eta^{\lambda\sigma}\varphi_{\mu\nu\lambda\sigma\mu_5\cdots \mu_s}=0,\quad
F_{\mu_1\cdots \mu_{s-2}}=\frac{1}{2}\eta^{\mu\nu}\varphi_{\mu\nu\mu_1\cdots \mu_{s-2}},\\
\label{v1}
&&
\alpha_{\mu_1\cdots \mu_{s-3}}=0, \quad \lambda_{(2)}^{\mu_1\cdots \mu_{s-2}}=0,\quad
\lambda_{(4)}^{\mu_1\cdots \mu_{s-4}}=0,\quad
\eta^{\mu\nu} \xi_{\mu\nu\mu_1\cdots \mu_{s-3}}=0.
\eeq
In this limit the action $S_0[\varphi, F, \alpha, \lambda_{(2)}, \lambda_{(4)}]$
(\ref{b1}) reduces to the Fronsdal action \cite{Fronsdal-1},
\beq
\nonumber
&&S_0[\varphi]=\int dx \Big[\varphi_{\mu_1\cdots\mu_s}\Box
\varphi^{\mu_1\cdots\mu_s}-
\frac{s(s-1)}{2}\eta^{\mu\nu}\eta_{\rho\sigma}\varphi_{\mu\nu\mu_3\cdots\mu_s}
\Box \varphi^{\rho\sigma\mu_3\cdots\mu_s}-\\
\nonumber
&&\qquad\qquad\qquad
-s\varphi_{\mu\mu_2\cdots\mu_s}\pa^\mu
\pa_\nu\varphi^{\nu\mu_2\cdots\mu_s}+
s(s-1)\eta^{\mu\nu}\varphi_{\mu\nu\mu_3\cdots\mu_s}
\pa_{\rho}\pa_\sigma\varphi^{\rho\sigma\mu_3\cdots\mu_s}-\\
\label{v2}
&&\qquad\qquad\qquad
-\frac{s(s-1)(s-2)}{4}\eta^{\mu_1\mu_2}
\varphi_{\mu_1\mu_2\mu_3\mu_4\cdots\mu_s}
\pa^{\mu_3}\pa_{\nu_1}\eta_{\nu_2\nu_3}
\varphi^{\nu_1\nu_2\nu_3\mu_4\cdots\mu_s}\Big]
\eeq
In its turn, the action  $S_{1\;loc}[\varphi,\phi,F]$ is transformed into
$S_{1\;loc}[\varphi,\phi]$,
\beq
\nonumber
&&S_{1\;loc}[\varphi,\phi]=2s\int dx\varphi_{\mu_{1}\mu_2\cdots \mu_{s}}\Big[
\pa^{\mu_1}\pa^{\mu_2}R_0^{\mu_{3}\cdots \mu_{s}}(\phi)-
\eta^{\mu_1\mu_2}\Box R_0^{\mu_{3}\cdots \mu_{s}}(\phi)-\\
\label{v3}
&&\qquad\qquad\qquad\qquad\qquad\qquad\quad-
\frac{(s-2)}{2}\eta^{\mu_1\mu_2}
\pa_{\rho}\pa^{\mu_3}R_0^{\mu_{4}\cdots \mu_{s}\rho}(\phi)
\Big],
\eeq
where functions $R_0^{\mu_{1}\cdots \mu_{s-2}}(\phi)$ are constructed from
$R^{\mu_{1}\cdots \mu_{s-2}}(\phi)$ in the limit $\eta_{\mu\nu}\rightarrow 0$
that corresponds to use traceless condition on gauge parameters,
$\eta_{\mu\nu}\xi^{\mu\nu\mu_1\cdots \mu_{s-3}}=0$, in the equations
(\ref{c16}), (\ref{c16a}).
Integrating by parts in the integral (\ref{v3}) allows the equivalent presentation for
action $S_{1\;loc}[\varphi,\phi]$
\beq
\nonumber
&&S_{1\;loc}[\varphi,\phi]=2s\int dx\Big[\pa^{\mu_1}\pa^{\mu_2}
\varphi_{\mu_{1}\mu_2\mu_3\cdots \mu_{s}}-
\eta^{\mu_1\mu_2}\Box \varphi_{\mu_{1}\mu_2\mu_3\cdots \mu_{s}}-\\
\label{v4}
&&\qquad\qquad\qquad\qquad\qquad-
\frac{(s-2)}{2}\eta^{\lambda\sigma}\pa^{\rho}\pa_{\mu_3}
\varphi_{\lambda\sigma\rho\mu_{4}\cdots \mu_{s}}
\Big]R_0^{\mu_{3}\cdots \mu_{s}}(\phi),
\eeq

The action $S[\varphi,\phi]=S_0[\varphi]+S_{1\; loc}[\varphi,\phi]$
describes in the explicit and closed
form with the help of tensor $R_0^{\mu_{1}\cdots \mu_{s-2}}(\phi)$ a consistent gauge
model invariant under  gauge transformations
\beq
\delta\varphi_{\mu_1\cdots\mu_s}=\pa_{(\mu_1}\xi_{\varphi_{\mu_2\cdots\mu_s})},
\quad \delta\phi=0
\eeq
subjected to traceless conditions
\beq
\eta^{\mu\nu}\eta^{\lambda\sigma}\varphi_{\mu\nu\lambda\sigma\mu_5\cdots \mu_s}=0, \quad
\eta^{\mu\nu}\xi_{\mu\nu\mu_{1}\cdots \mu_{s-3}}=0
\eeq
of the Fronsdal theory. In what follows we will use the notation ${\it "partial \;\;on-shell"}$
for the result of eliminating auxiliary fields from the deformed unconstrained action.

\section{Special cases of vertices}

It is remarkable fact that
any gauge-invariant interaction between massless spin
$s$ field $\varphi$
and any number of scalar fields $\phi$ is described by one $(s-2)$-rank tensor
$R^{\mu_1\cdots \mu_{s-2}}(\phi)$ in a simple enough way (\ref{c18}) using
the unconstrained formulation \cite{BGK} for new method \cite{BL-1}. Eliminating auxiliary
fields leads to the similar situation when mentioned above interaction
is defined by the tensor
$R_0^{\mu_3\cdots \mu_s}(\phi)$ in the form (\ref{v3}).
As a general result, for gauge theories under consideration solutions
to the problem of suitable deformations
leading to local gauge-invariant interactions have an  unique form.
Modeling an explicit form of local tensors $R^{\mu_1\cdots \mu_{s-2}}(\phi)$ we arrive at
the explicit structure of vertices due to (\ref{c18}). These vertices will be automatically
invariant under original gauge transformations. Then we can reproduce the corresponding results
to the partial on-shell due to (\ref{v4}).
Below we are going to demonstrate some special
cases of vertices obtained in this way.

\subsection{Cubic vertices $\sim\varphi\phi\phi$ for arbitrary spin $s$}

We begin construction of consistent interactions  for the  case of cubic vertices
because they belong to the simplest class of interactions in gauge theories. It is clear
that function $R^{\mu_3\cdots \mu_s}(\phi)$ should be at least quadratic in field $\phi$
and we restrict ourself by the quadratic dependence in scalar field of this function.
To reproduce tensor structure of $R^{\mu_3\cdots \mu_s}(\phi)$ we have to use partial
derivatives $\pa^{\mu}$ and the metric tensor $\eta_{\mu\nu}$ of Minkowski space-time.
The minimum number of partial derivatives cannot be less than $s-2$. Let us use
the minimal number of partial derivatives in construction of $R^{\mu_3\cdots \mu_s}(\phi)$.
Then, the more general form of $R^{\mu_3\cdots \mu_s}(\phi)$ reads
\beq
\nonumber
R^{\mu_3\cdots \mu_s}(\phi)&=&c_0 \phi\;\pa^{\mu_3}\cdots \pa^{\mu_s}\!\phi+
\sum_{k=1}^{m_s} c_k \pa^{(\mu_3}\cdots \pa^{\mu_{k+2}}\phi\;\pa^{\mu_{k+3}}
\cdots \pa^{\mu_s)}\phi+\\
\nonumber
&&+d_0\phi\; \eta^{(\mu_3\mu_4}\Box \pa^{\mu_5}\cdots \pa^{\mu_s)}\phi+
d_1\pa_{\rho}\phi\;\eta^{(\mu_3\mu_4}\pa^{\mu_5}\cdots \pa^{\mu_s)\rho}\phi+\\
\label{f1}
&&+\sum_{k=2}^{m_s} d_k\eta^{(\mu_3\mu_4}\Box\pa^{\mu_5}\cdots \pa^{\mu_{k+2}}\phi\;
\pa^{\mu_{k+3}}\cdots \pa^{\mu_s)}\phi ,
\eeq
where $c_i,\;d_i \;(i=0,1,...,m_s)$ are constants having coinciding dimensions,
${\rm dim}(c_0)={\rm dim}(d_0)={\rm dim}(c_i)={\rm dim}(d_i),\;( i=1,2,...,m_s)$.
In Eq. (\ref{f1}) the notation
\beq
\label{f2}
m_s=\left[\frac{s}{2}\right] -\frac{1+(-1)^s}{2}
\eeq
is used. Here $[a]$ denotes the integer part of number $a$.
Constructed cubic vertices (\ref{c18}) with the help of (\ref{f1}) form a $(2m_s+1)$-
parameter family of interactions between integer spin $s$ and scalar fields
when vertices contain $s$ derivatives.
Notice that on this stage there are cubic vertices involving
auxiliary fields as well, $\sim F\phi\phi$.

Of special interest is related with consideration of cubic vertices on the partial on-shell
 because usually description of interactions with higher spin fields is performed
in terms of the double trace restrictions. They are described by action (\ref{v3}) with using
the function
\beq
R_0^{\mu_3\cdots \mu_s}(\phi)=c_0 \phi\;\pa^{\mu_3}\cdots \pa^{\mu_s}\phi+
\sum_{k=1}^{m_s} c_k \pa^{(\mu_3}\cdots \pa^{\mu_{k+2}}\phi\;\pa^{\mu_{k+3}}
\cdots \pa^{\mu_s)}\phi .
\eeq
These vertices $\sim\varphi\phi\phi$ belong to $m_s$ -parameter family
of interactions with minimal number of
derivatives equal to $s$.

\subsection{Vertices of any order for massless spin 4 field}
Minimal value of spin when one needs to use the unconstrained formulation
for higher spin fields to apply the new method in constructing interactions is equal to $s=4$.
The more general form of function $R^{\mu\nu}(\phi)$ responsible for cubic interactions reads
\beq
\label{r1}
R^{\mu\nu}(\phi)=c_1\pa^{\mu}\pa^{\nu}\!\phi\;\phi+c_2\pa^{\mu}\phi\;\pa^{\nu}\phi+
d_1\eta^{\mu\nu}\Box \phi\;\phi+d_2\eta^{\mu\nu}\pa_{\rho}\phi\;\pa^{\rho}\phi .
\eeq
In this case the action (\ref{c18}) can be written in the form
\beq
\label{r2}
S_{1\;loc}[\varphi,\phi,F]=8\int dx\Big[\pa^{\lambda}\pa^{\sigma}\!
\varphi_{\mu\nu\lambda\sigma}-
2\Box F_{\mu\nu} -
2\pa^{\rho}\pa_{\mu}F_{\rho\nu}
\Big] R^{\mu\nu}(\phi).
\eeq
This action belongs to a three-parameter family of interactions between massless spin 4
and scalar fields.
Passing to the partial on-shell is performed with the help of function
\beq
\label{r3}
R_0^{\mu\nu}(\phi)=c_1\pa^{\mu}\pa^{\nu}\phi\;\phi+c_2\pa^{\mu}\phi\;\pa^{\nu}\phi
\eeq
with the following result for action (\ref{v3})
\beq
\label{r4}
S_{1\;loc}[\varphi,\phi]=8\int dx\Big[\pa^{\lambda}\pa^{\sigma}
\varphi_{\mu\nu\lambda\sigma}
-
\eta^{\lambda\sigma}\Box \varphi_{\mu\nu\lambda\sigma}-
\eta^{\lambda\sigma}
\pa_{\mu}\pa^{\rho}\varphi_{\nu\lambda\sigma\rho}
\Big]R_0^{\mu\nu}(\phi),
\eeq
In its turn this action belongs to  one-parameter family of interactions.
This confirms the result obtained in \cite{L-s4-3}.

It is not so hard to extend the obtained result for vertices of any order in scalar fields.
To do this it is enough to choice the function $R_{(n)}^{\mu\nu}(\phi)$ responsible for desired
interactions in the form
\beq
\label{r5}
R_{(n)}^{\mu\nu}(\phi)=R^{\mu\nu}(\phi)\;\phi^{n-2}, \quad n=3,4,...,
\eeq
where $R^{\mu\nu}(\phi)$ is defined in (\ref{r1}).
The corresponding action is
\beq
\label{r6}
S^{(n)}_{1\;loc}[\varphi,\phi,F]=8\int dx\Big[\pa^{\lambda}\pa^{\sigma}\!
\varphi_{\mu\nu\lambda\sigma}-
2\Box F_{\mu\nu} -
2\pa^{\rho}\pa_{\mu}F_{\rho\nu}
\Big] R^{\mu\nu}(\phi)\;\phi^{n-2},
\eeq
depending on three independent parameters. Transition to the partial on-shell is created
by the function $R_{0\;(n)}^{\mu\nu}(\phi)$,
\beq
\label{r7}
R_{0\;(n)}^{\mu\nu}(\phi)=R^{\mu\nu}_0(\phi)\;\phi^{n-2}, \quad n=3,4,...,
\eeq
leading to the action
\beq
\label{r8}
S^{(n)}_{1\;loc}[\varphi,\phi]=8\int dx\Big[\pa^{\lambda}\pa^{\sigma}
\varphi_{\mu\nu\lambda\sigma}
-
\eta^{\lambda\sigma}\Box \varphi_{\mu\nu\lambda\sigma}-
\eta^{\lambda\sigma}
\pa_{\mu}\pa^{\rho}\varphi_{\nu\lambda\sigma\rho}
\Big]R_0^{\mu\nu}(\phi)\;\phi^{n-2}.
\eeq
The action (\ref{r8}) depends on one independent parameter.

\section{Discussion}

In the present paper studies of models which include  high spin fields interacting with
matter fields and admit explicit and closed descriptions have been continued.
Construction of these models is based on applications of the new method
\cite{BL-1,BL-2,L-2022} to the deformation procedure for gauge theories \cite{BH,H}.
First examples were related to  massless spin 3 fields when the original gauge actions
belonged to the class of first-stage  reducible theories
in the terminology of BV-formalism \cite{BV,BV1}. On this way it was proved that cubic
vertices for interacting  spin 3 field with scalar and massive vector field being invariant
under original gauge transformations were forbidden
while quartic vertices can be explicitly constructed \cite{L-ejpc-s3,L-22-2}.
Later on, for the first time in the history of the high spin theory quintic
vertices describing
local interactions between  spin 3 field and scalar and vector fields in exact form
were found \cite{L-s5}. All found vertices with massless spin 3 field
\cite{L-ejpc-s3,L-22-2,L-s5} had an unique form of local functionals which were invariant under
original gauge transformations. Next step in constructions of consistent interactions
within the new method was made in \cite{L-s4-3} where the case of massless spin 4 field
interacting with scalar field was studied. It was found that local non-trivial cubic vertices
invariant under original gauge transformations in both unconstrained and constrained
settings can be constructed but in comparison with the spin 3 fields the interactions
contained arbitrariness. In more interesting constrained (without auxiliary fields)
description this arbitrariness was defined by one parameter. This conclusion
was supported in Sec. 5 of the present study  when it was proved that vertices
of any order for interactions of one massless
spin 4 field with any number of scalar fields in constrained description
depend on one parameter.

Main efforts of the present studies were concentrated on the derivation
of general conditions allowing gauge invariant interactions of massless integer spin
$s$ fields with scalar field. It was required to use the unconstrained formulation
\cite{BGK}
of the Fronsdal theory \cite{Fronsdal-1} in the framework of new method \cite{BL-1}.
We considered a special form of anticanonical transformations which allowed us
to arrive at two
important properties: a) the deformations of the original gauge action  will be nontrivial,
b) the original gauge symmetry will be preserved during the deformations.
It was proved that the deformation of local part of deformed action was ruled by one
$(s-2)$-rank tensor of scalar fields leading to simple enough explicit form of vertices
(\ref{c19}), (\ref{c21}). In turn, eliminating all auxiliary fields led to the
formulation of the vertices in terms of double traceless spin fields of the Fronsdal theory
(\ref{v3}), (\ref{v4}).
General results were illustrated by two examples. The first one was devoted to
explicit construction
of cubic vertices which describe  interacting massless integer spin $s$ fields with scalar
field using assumption about minimal number (equal to $s$)
of partial derivatives in vertices.
Existing arbitrariness in the constrained description was described by $m_s$ constants
(\ref{f2}). With an increase in the value of the spin, the arbitrariness in vertices
also proportionally increased.
The second example presented construction of vertices of any order describing interactions
of one massless spin 4 field with any number of scalar fields. Constrained presentation
of these vertices  used one parameter only.
Here, it is interesting to note that after construction of models with quintic vertices
\cite{L-s5}, in the high spin theory local gauge invariant
vertices of any order are introduced.

Results obtained in the present study can be easily extended
to more general models including
additional fields, for example, massive vector field $A^{\mu}$ when the initial theory
(\ref{c1}) can be extended by the action
\beq
\label{c100}
S_0[A]=-\int dx \Big(\frac{1}{4}F_{\mu\nu}F^{\mu\nu}+
\frac{1}{2}m_0^2A_{\mu}A^{\mu}\Big),\quad
F_{\mu\nu}=\pa_{\mu}A_{\nu}-\pa_{\nu}A_{\mu}
\eeq
for free massive field $A^{\mu}$, and the function $R^{\mu_3\cdots \mu_s}$ defining special
anticanonical transformations will depend on fields $\phi$ and $A^{\mu}$. For example,
local vertices describing interactions of one integer spin $s$ field with scalar field and
$s-4$ massive vector fields are generated with the help of the following function
\beq
\nonumber
&&R^{\mu_1\cdots \mu_{s-2}}(A,\phi)=
c_1\pa^{\mu_1}\pa^{\mu_2}\!A^{\mu_3\;}A^{\mu_4}\cdots A^{\mu_{s-2}}\;\phi+
c_2\pa^{\mu_1}\!A^{\mu_2}\;\pa^{\mu_3}\!A^{\mu_4\;}A^{\mu_5}\cdots A^{\mu_{s-2}}\;\phi+\\
\nonumber
&&\qquad+
c_3\pa^{\mu_1}\!A^{\mu_2}\;A^{\mu_3\;}A^{\mu_5}\cdots A^{\mu_{s-3}}\;\pa^{\mu_{s-2}}\phi +
c_4A^{\mu_1}\;A^{\mu_2\;}A^{\mu_5}\cdots A^{\mu_{s-4}}\;
\pa^{\mu_{s-3}}\pa^{\mu_{s-2}}\phi +\\
\nonumber
&&\qquad+d_1\eta^{\mu_1\mu_2}\Box\!A^{\mu_3}\;A^{\mu_4}\cdots A^{\mu_{s-2}}\;\phi+
d_2\eta^{\mu_1\mu_2}\pa^{\rho}A^{\mu_3}\;\pa_{\rho}\!A^{\mu_4}\;A^{\mu_5}
\cdots A^{\mu_{s-2}}\;\phi+\\
\label{d10}
&&\qquad+d_3\eta^{\mu_1\mu_2}A^{\mu_3}\;
A^{\mu_4}\cdots
A^{\mu_{s-2}}\;\Box\phi+
d_4\eta^{\mu_1\mu_2}\pa_{\rho}\!A^{\mu_1}\;A^{\mu_2}\cdots
A^{\mu_{s-4}}\eta^{\mu_{s-3}\mu_{s-2}}\;\pa^{\rho}\!\phi .
\eeq
In fact, there exist a seven-parameter family of vertices invariant
under original gauge transformations.
Passing in (\ref{d10})
to the partial on-shell leads to the function
\beq
\nonumber
&&R_0^{\mu_1\cdots \mu_{s-2}}(A,\phi)=
c_1\pa^{\mu_1}\pa^{\mu_2}\!A^{\mu_3\;}A^{\mu_4}\cdots A^{\mu_{s-2}}\;\phi+
c_2\pa^{\mu_1}\!A^{\mu_2}\;\pa^{\mu_3}\!A^{\mu_4\;}A^{\mu_5}\cdots A^{\mu_{s-2}}\;\phi+\\
\label{d11}
&&\qquad+
c_3\pa^{\mu_1}\!A^{\mu_2}\;A^{\mu_3\;}A^{\mu_5}\cdots A^{\mu_{s-3}}\;\pa^{\mu_{s-2}}\phi +
c_4A^{\mu_1}\;A^{\mu_2}\cdots A^{\mu_{s-4}}\;
\pa^{\mu_{s-3}}\pa^{\mu_{s-2}}\phi
\eeq
responsible for a three-parameter family of local vertices.

Finally, we can state that  all vertices containing one massless integer spin
field  interacting with  any number of massive vector and scalar fields admit
a closed and
exact description in terms of local functionals invariant
under original gauge transformations.

\section*{Acknowledgments}
\noindent
The author thanks K. Koutrolikos and Yu.M. Zinoviev for correspondence and useful
discussions.
The work is supported by the Ministry of
Education of the Russian Federation, project FEWF-2020-0003.

\end{document}